\documentclass[prl,superscriptaddress,floatfix,preprintnumbers,twocolumn]{revtex4}

\usepackage{url,color}
\usepackage{graphicx}
\usepackage{bm} 
\usepackage{amsfonts,amsmath,amssymb,amsbsy,latexsym}

\newcommand{\bc}{\begin{center}}
\newcommand{\ec}{\end{center}}
\newcommand{\be}{\begin{equation}}
\newcommand{\ee}{\end{equation}}
\newcommand{\ba}{\begin{eqnarray*}}
\newcommand{\ea}{\end{eqnarray*}}
\newcommand{\bna}{\begin{eqnarray}}
\newcommand{\ena}{\end{eqnarray}}

\newcommand{\mpaa}{\begin{minipage}[t]{7.5cm}}
\newcommand{\mpea}{\end{minipage}}

\begin{document}

\title{Normal and Anomalous Diffusion in Soft Lorentz Gases}

\title{Normal and Anomalous Diffusion in Soft Lorentz Gases}

\author{Rainer Klages}
\email{r.klages@qmul.ac.uk}
\affiliation{Queen Mary University of London, School of Mathematical Sciences, Mile End Road, London E1 4NS, UK}
\affiliation{Institut f\"ur Theoretische Physik, Technische Universit\"at Berlin, Hardenbergstra{\ss}e 36, 10623 Berlin, Germany}
\affiliation{Institute for Theoretical Physics, University of Cologne, Z\"ulpicher Stra{\ss}e 77, 50937 Cologne, Germany}
\author{Sol Selene Gil Gallegos}
\affiliation{Queen Mary University of London, School of Mathematical Sciences, Mile End Road, London E1 4NS, UK}
\author{Janne Solanp\"a\"a}
\affiliation{Computational Physics Laboratory, Tampere University, P.O. Box 692, FI-33014 Tampere, Finland}
\author{Mika Sarvilahti}
\affiliation{Computational Physics Laboratory, Tampere University, P.O. Box 692, FI-33014 Tampere, Finland}
\author{Esa R\"as\"anen}
\affiliation{Computational Physics Laboratory, Tampere University, P.O. Box 692, FI-33014 Tampere, Finland}

\date{\today}


\begin{abstract}
  Motivated by electronic transport in graphene-like structures, we
  study the diffusion of a classical point particle in Fermi
  potentials situated on a triangular lattice. We call this system a
  soft Lorentz gas, as the hard disks in the conventional periodic
  Lorentz gas are replaced by soft repulsive scatterers. A thorough
  computational analysis yields both normal and anomalous (super)
  diffusion with an extreme sensitivity on model parameters. This is
  due to an intricate interplay between trapped and ballistic periodic
  orbits, whose existence is characterized by tongue-like structures
  in parameter space. These results hold even for small softness
  showing that diffusion in the paradigmatic hard Lorentz gas is not
  robust for realistic potentials, where we find an entirely
  different type of diffusion.
\end{abstract}

\maketitle

The rise of new micromanipulation techniques, molecular nanodevices
and nanotechnologies has fuelled the scientific interest in {\em small
  systems} \cite{BLR05,Rit08,RRS10,KJJ13}. These are objects composed
of small numbers of particles far from the thermodynamic limit, which
exhibit only a few relevant degrees of freedom \cite{KJJ13}. Their
microscopic equations of motion are typically highly nonlinear
yielding fluctuations with macroscopic statistical properties
reminiscent of interacting many-particle systems. Small systems can
thus serve as a laboratory for understanding the emergence of
irreversibility and complexity from chaotic dynamics
\cite{BaPo97,CFLV08}. They become especially interesting under
nonequilibrium conditions, where they exhibit macroscopic transport
phenomena like diffusion.  By combining nonlinear dynamics with
nonequilibrium statistical physics the origin of macroscopic transport
from microscopic chaos in small systems was explained by formulas
expressing transport coefficients in terms of dynamical systems
quantities \cite{Gasp,Do99,Kla06,EvMo08}. Similarly irreversible
entropy production was found to emerge from fractal measures
\cite{Gasp,Do99} and fractal attractors
\cite{Rue99,EvMo08,Hoo15}. These results paved the way for fundamental
concepts like the chaotic hypothesis generalising Boltzmann's ergodic
hypothesis \cite{GaCo95b} and fluctuation theorems generalising the
second law of thermodynamics \cite{Do99,Kla06,KJJ13,ESW16}.

Classical transport in small systems has a quantum mechanical analogue
as electronic transport in solid-state nanodevices \cite{Mass08}.
Recently growing interest has been attracted by periodic nanosystems
such as {\em artificial graphene} \cite{PGLMP13} fabricated in
semiconductor heterostructures \cite{Gibertini,Rasanen,Wang} or on
metallic surfaces \cite{GMKGM12,PRNAR16}. In the latter case, the
electrons are confined to a honeycomb geometry by CO molecules
positioned with a scanning tunneling microscope in a triangular
configuration.  This system exhibits the properties of graphene but in
a setup that is tunable regarding, e.g., the electronic density,
lattice constant, geometry, and the coupling with the enviroment.

Interestingly, the topology of ``molecular graphene'' as described
above is exactly the same as one of the most paradigmatic models in
dynamical systems theory, the periodic {\em Lorentz gas}
\cite{Lo05,Gasp,Do99,Kla06,Sza00,Dett14}.  Lorentz gases mimick the
motion of classical electrons in metals.  They consist of a point
particle scattering elastically with fixed {\em hard} spheres
distributed either randomly or periodically in space. Originally they
were devised to reproduce Drude's theory from microscopic dynamics
\cite{Lo05}. In groundbreaking mathematical works Lorentz gases were
shown to exhibit chaos and well-behaved transport properties
\cite{BuSi80a,BuSi80b}, followed by understanding diffusion in
computer simulations combined with stochastic theory
\cite{MaZw83,ZGNR86,BoDe86}. Lorentz gases thus became standard models
to explain the interplay between chaos and transport: Highlights were
a proof of Ohm's law from first principles \cite{Ch1}, the analytical
and numerical calculation of Lyapunov exponents \cite{Gasp,Do99,Hoo15}
and fractal attractors \cite{Hoo15}, as well as developing a chaotic
scattering theory of transport \cite{Gasp}. The growing interest in
graphene-like systems now brings direct technological relevance to
investigate classical diffusion in {\em soft} Lorentz gases equipped
with more realistic potentials.

\begin{figure}[t]
\includegraphics[width=\linewidth]{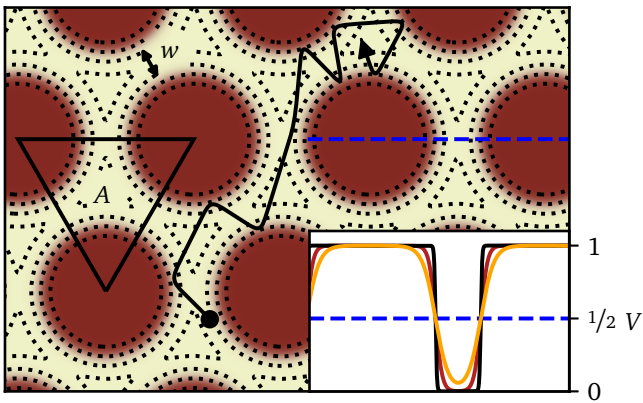}
\caption{The soft Lorentz gas: A point particle moves in a plane of
  partially overlapping Fermi potentials (inset) whose centres are
  situated on a triangular lattice (main figure). The dotted lines are
  contour lines, $w$ denotes the minimal distance between adjacent
  potentials for total energy $E=1/2$, and $A$ defines a triangular
  unit cell. The inset shows Fermi potentials along the dashed (blue)
  line in the main part} for different values of the softness
parameter $\sigma$ defined in Eq.~(\ref{eq:fp}).
\label{fig:model}
\end{figure}

The conventional two-dimensional periodic Lorentz gas is a Hamiltonian
particle billiard in which a point particle of mass $m$ performs free
flights with constant velocity $v$ between elastic collisions at hard
disks of radius $r_0$. The centers of these disks form the nodes of a
triangular lattice with lattice spacing $2r_0+w$, where $w$ denotes
the smallest distance between two nearby disks. Here, following
previous studies on artificial graphene \cite{aichinger}, we introduce
a {\em soft Lorentz gas}, where the hard disks are replaced by Fermi
potentials, \be
V(\mathbf{r})=\frac{1}{1+\exp\Big((|\mathbf{r}|-r_o)/\sigma\Big)}
\label{eq:fp}\quad , \ee with $\sigma$ determining the softness of the
potential; see Fig.~\ref{fig:model}. Related models have been used to
reproduce experimental results on the magnetoresistance of electrons
in semiconductor antidot lattices
\cite{Geis90,LKP91,Weis91,FGK92,FlSS}.  In the following we set
$m=r_0=1$ by keeping the total energy constant, $E=1/2$. We thus have
two control parameters, $\sigma$ and the minimal gap size $w$ between
two nearby potentials for the given energy $E$. Making $\sigma$
smaller we approach the hard scatterer limit of the conventional
Lorentz gas. A crucial question is to which extent chaotic diffusion
in the hard Lorentz gas \cite{Gasp,Do99,Sza00,Dett14,Ch1,Kla06,Lo05}
is robust by softening the potential, i.e., for more realistic
models. In this Letter we show that even a slight softening introduces
substantial additional complexity leading to entirely new transport
properties.

Our key quantity is the diffusion coefficient 
\be 
D=\lim_{t \rightarrow \infty}
\frac{\langle(\mathbf{r}(t)-\mathbf{r}(0))^2\rangle}{4t}\quad ,
\label{eq:dc} 
\ee 
where the numerator denotes the mean square displacement (MSD) for the
position $\mathbf{r}(t)$ of a particle at time $t$. The angular
brackets hold for an ensemble average over initial conditions.  If
the MSD grows linearly in time, the above limit exists and the system
exhibits normal diffusion. If the MSD grows faster than linear in time
this limit diverges, and the system displays superdiffusion
\cite{KRS08}. Technical details of the simulations carried out with
the Bill2D software package \cite{bill2d} are explained in Sec.~1 of
our Supplemental Material \cite{suppl}, which includes 
Ref.~\cite{yoshida_4th_order}.

\begin{figure}[t]
\includegraphics[width=\linewidth]{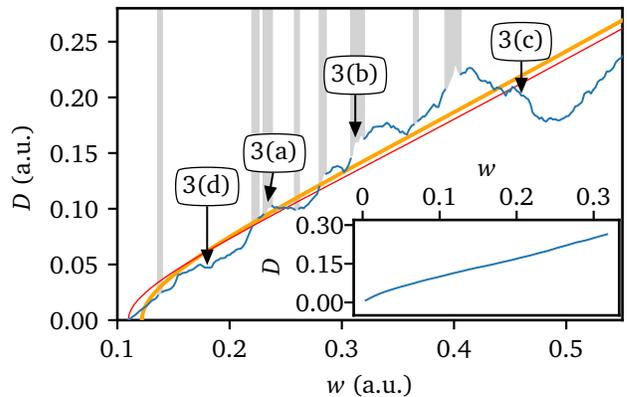}
\caption{Diffusion coefficient $D$ as a function of the gap size $w$.
  The (blue) wiggled line shows simulation results for $D(w)$
  Eq.~(\ref{eq:dc}) in a slightly softened potential ($\sigma=0.05$ in
  Eq.~(\ref{eq:fp})). The thick (orange) line represents the
  corresponding analytical random walk approximation $D_B$, the thin
  (red) line the numerical $D_{B,num}$ as explained in the text.  The
  labeled numbers 3(a) to (d) refer to the periodic orbits depicted in
  Fig.~\ref{fig:poi}. Grey columns indicate parameter intervals in
  which $D(w)$ does not exist. The inset displays $D(w)$ obtained from
  simulations for the conventional hard Lorentz gas \cite{KlDe00}.}
\label{fig:diffw}
\end{figure}

Figure~\ref{fig:diffw} depicts the diffusion coefficient $D$ as a
function of the gap size $w$ between the scattering centers for a
slightly softened (main part) and the hard (inset) Lorentz gas.  While
for the hard scatterers $D(w)$ is monotonically increasing and looks
rather smooth, in the soft model $D$ is a non-monotonic, highly
complicated function of $w$. This suggests that the diffusive
properties must have changed profoundly. The diffusion coefficient for
the hard Lorentz gas has been analysed in detail in previous
literature, cf.\ Sec.~3.A in \cite{suppl}, which includes
Refs.~\cite{Sin70,GilSan09,AnMo12,CEG91,CGLS14,CGLS14c,ZPFDB18,LiLi92}.
Here we first explore whether there is any simple diffusion law for
the soft model revealing an at least on average monotonic increase of
$D(w)$ by ignoring any fine structure.  We find that a Boltzmann-type
random walk approximation works well to understand the coarse
functional form of $D(w)$ \cite{KlDe00}. For this we assume that
diffusion is governed by `flights' of length $\ell_c$ during time
intervals $\tau_c$ after which a particle experiences a `collision'.
We define a collision as an event where a particle hits the contour
line of a scatterer at $E=1/2$ in the triangular unit cell $A$
displayed in Fig.~\ref{fig:model}. By assuming in the spirit of
Boltzmann's molecular chaos hypothesis that all collisions are
uncorrelated, the diffusion coefficient can be approximated as
$D_B(w)=\ell_c^2(w)/(4\tau_c(w))$. In Sec.~2 of \cite{suppl} we derive
an analytical formula for $D_B$ as well as an improved numerical
version $D_{B,num}$. The results are shown as a pair of lines in
Fig.~\ref{fig:diffw}: Both yield an approximately linear increase of
$D$ for larger $w$, which matches well to the coarse functional form
of the simulation results. For smaller $w$ our analytical aproximation
does not reproduce the onset of diffusion correctly while our improved
numerical version captures it at least qualitatively well.

We now focus on the pronounced irregular fine structure of $D(w)$ in
the soft system, which is in sharp contrast to the diffusion
coefficient of the hard disk model. Irregular diffusion coefficients
have been reported for parameter-dependent deterministic diffusion in
much simpler chaotic dynamical systems, such as one-dimensional maps
\cite{RKD,GrKl02,KoKl02,KoKl03}, the standard map \cite{MaRo14,HaCo18}
and particle billiards \cite{KlDe00,HaGa01,HaKlGa02,MaKl03}. To our
knowledge this is the first time that a diffusive fine structure has
been unambiguously revealed in quite a realistic soft Hamiltonian
system. For the hard Lorentz gas irregularities in $D(w)$ also exist
but are extremely tiny \cite{KlDe00,KlKo02}, hence barely visibly in
Fig.~\ref{fig:diffw}. A second crucial difference is that our softened
model generates an intricate set of superdiffusive parameter regions
in which $D(w)$ does not exist.  The hard Lorentz gas displays only
superdiffusion for all parameters $w> w_{\infty}$ after a specific
geometric transition at $w_{\infty}\simeq0.3094$ \cite{Dett12,FVG17}
by exhibiting superdiffusion that is different from the soft model as
discussed in Sec.~3.A of \cite{suppl}.

\begin{figure}[t]
\includegraphics[width=\linewidth]{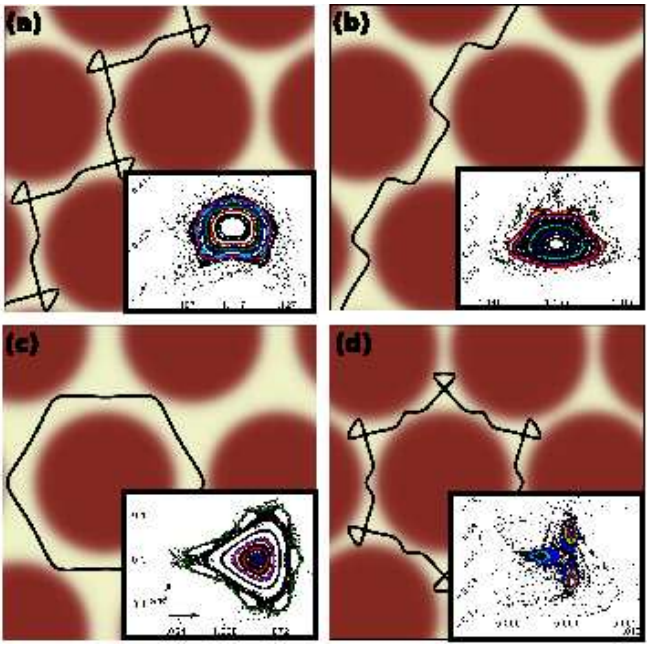}
\caption{Periodic orbits and islands of periodicity in phase space at
  different parameter values $w$ corresponding to
  Fig.~\ref{fig:diffw}. Shown in position space are characteristic
  periodic orbits for (a) $w=0.234$, (b) $w=0.31$, (c) $w=0.46$, (d)
  $w=0.18$. (a) and (b) feature quasi-ballistically propagating orbits
  yielding superdiffusive parameter regions in Fig.~\ref{fig:diffw}
  while (c) and (d) generate local minima in $D(w)$. The insets
  display associated islands of periodicity in the Poincar\'e surface
  of section phase space $(x,\sin\theta)$ as defined in the text.}
\label{fig:poi}
\end{figure}

The origin of the anomalous diffusion as well as of the irregularities
in $D(w)$ of the soft Lorentz gas can be understood in terms of
periodic orbits \cite{CvGS92,CAMTV01,KlKo02,KoKl02,KoKl03}, as is
explained by Fig.~\ref{fig:poi}. It shows orbits both in position
space and insets of corresponding Poincar\'e surfaces of section at
four specific parameter values of $w$: Figs.~\ref{fig:poi} (a) and (b)
refer to quasi-ballistically propagating periodic orbits while (c) and
(d) represent localised ones. These periodic orbits exhibit different
structures due to different types of scattering, as is reflected in
the corresponding Poincar\'e surfaces of section. The variables
$(x,\sin\theta)$ for the latter are defined on the boundary where a
particle leaves the unit cell $A$ in Fig.~\ref{fig:model}.  Here $x$
represents the position of the particle in a gap, $\sin\theta$ is the
angle between its velocity vector and the normal to the boundary.
These islands of periodicity are typically extremely small and very
difficult to detect in the whole phase space.  By matching the
parameter values of $w$ for these periodic orbits to the structure of
$D(w)$ in Fig.~\ref{fig:diffw} we see that the two propagating orbits
correspond to two superdiffusive regions while the two localised
orbits identify (approximately) two local minima in the curve.  While
localised orbits only slightly suppress normal diffusion without
making it anomalous \cite{RKD,KlDo99,KlKo02}, islands of periodicity
in phase space, also called {\em accelerator modes}
\cite{MaRo14,HaCo18}, generate superdiffusion
\cite{GZR87,GZR88,Zas02,KRS08}. A more detailed analysis yields that
all these periodic orbits are topologically extremely unstable under
parameter variation: They exhibit complicated bifurcation scenarios
that eventually destroy any superdiffusive window leading to parameter
regions of normal diffusion before new bifurcations create new
superdiffusive windows \cite{Gall18}.

\begin{figure}[t]
\includegraphics[width=\linewidth]{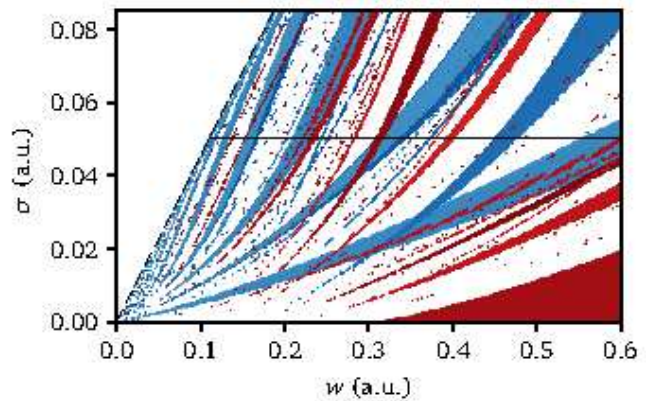}
\caption{Regions of periodic orbits in the parameter space of gap size
  $w$ and potential softness parameter $\sigma$. Blue dots represent
  localised periodic orbits like (c), (d) in Fig.~\ref{fig:poi} while
  red dots correspond to quasi-ballistic orbits like (a), (b) therein.
  The black horizontal line at $\sigma=0.05$ yields a cut through the
  parameter space corresponding to the diffusion coefficient $D(w)$ in
  Fig.~\ref{fig:diffw}. }
\label{fig:tongues}
\end{figure}

Periodic orbits thus form the backbone to understand the complicated
structure of the parameter-dependent diffusion coefficient in
Fig.~\ref{fig:diffw}. We now explore them in the full parameter space
$(w,\sigma)$. For each point in $(w,\sigma)$, the numerical discovery
of a localised periodic orbit or a quasi-ballistic trajectory is
marked in Fig.~\ref{fig:tongues} as a blue or a red dot, respectively.
Interestingly, our chart reveals a very regular topological structure
underlying the seemingly totally irregular curve of $D(w)$ in
Fig.~\ref{fig:diffw}, which lives on the horizontal black line at
$\sigma=0.05$ in Fig.~\ref{fig:tongues}. We see that all periodic
orbits form regular `tongues' in parameter space which, however, we
could not fit with simple functional forms like exponential, stretched
exponential, or power laws. Whenever a tongue crosses the horizontal
black line at $\sigma=0.05$ we have a local extremum in the $D(w)$
curve of Fig.~\ref{fig:diffw}. Further details of this connection are
described in Ref.~\cite{Gall18}. In Sec.~3.B of \cite{suppl} we
explore the impact of these tongues on the diffusion coefficient under
variation of $\sigma$. Therein we see that on a coarse scale $D(w)$ of
the hard Lorentz gas is approached continuously by decreasing
$\sigma$, interrupted by superdiffusive regions due to quasi-ballistic
tongues. This scenario is in line with a mathematical theory on the
existence of elliptic islands in the phase space of closed,
non-diffusive billiards that are softened \cite{TRK98,RKT99}. In these
references the authors conjecture that islands are dense with respect
to Lebesgue measure in parameter space for small $\sigma$. If this
holds true one expects $D(w)$ to be an irregular curve on arbitrarily
fine scales with fractal properties
\cite{GrKl02,Kla06,KlKo02,RKD,KlDo99}.

In summary, we have studied diffusion under parameter variation in a
soft Lorentz gas, which we put forward as a model for electronic
transport in artificial graphene. We have found that the normal
diffusion observed in the paradigmatic Lorentz gas with hard
scatterers is not robust when softening them: Instead, the type of
diffusion immediately changes dramatically generating an entirely
different diffusion coefficient. This raises doubts about a universal
applicability of the standard Lorentz gas for describing transport in
realistic systems. In the soft Lorentz gas the diffusion coefficient
turns out to be a highly irregular function under variation of control
parameters with regions exhibiting superdiffusion. This is explained
in terms of periodic orbits that are topologically unstable under
parameter variation while exhibiting very regular structures in
parameter space.  Analogous results hold for varying the energy $E$ as
a parameter \cite{Gall18}, which experimentally corresponds to
changing the temperature of the system. Note that in superdiffusive
parameter regions ergodicity is broken, hence for single particle
experiments there will be a dependence on initial conditions
\cite{MaRo14,HaCo18}. In real systems with thermal noise we expect
these superdiffusive regions, ergodicity breaking and irregularities
on fine scales to disappear, however, larger irregularities should
persist under noise \cite{HaGa01,RKla01b}. Our results motivate to
construct a more rigorous theory for calculating the diffusion
coefficient curve in Fig.~\ref{fig:diffw} from first principles,
possibly based on generating partitions \cite{ChriPol95}, which will
be an extremely difficult task \cite{CAMTV01,Kla06}. A second crucial
challenge is to test for the diffusion coefficient of
Fig.~\ref{fig:diffw} in experiments.

\begin{acknowledgments}
  This work was supported by the Academy of Finland (grants
  no.\ 267686 and 304458).  JS, MS and ER acknowledge the Finnish IT
  Center for Science and the Tampere Center for Scientific Computing
  for computational resources. SSGG thanks the Mexican National
  Council for Science and Technology (CONACyT) for support by
  scholarship no.\ 262481. R.K.\ thanks Prof.~Krug (U.\ of Cologne),
  Klapp and Stark (TU Berlin) for hospitality and Profs.~Turaev,
  Rom-Kedar and Barkai for interesting discussions. He received
  funding from the Office of Naval Research Global and from the London
  Mathematical Laboratory, where he is an External Fellow.
\end{acknowledgments}


\newpage

\setcounter{equation}{0}
\setcounter{figure}{0}
\setcounter{table}{0}
\setcounter{page}{1}
\renewcommand{\figurename}{Supplementary Figure}
\renewcommand{\tablename}{Supplementary Table}
\renewcommand{\theequation}{S\arabic{equation}}
\renewcommand{\thetable}{S\arabic{table}}
\renewcommand{\thefigure}{S\arabic{figure}}

\begin{appendix}

\onecolumngrid

\centerline{\Large \bf Supplemental Material}

\section{1. Technical details of the numerical simulations}
\label{sec:numsim}

The numerical simulations were carried out with the \emph{bill2d}
software package~\cite{sbill2d} for classical dynamics. The
time-propagation was performed using the 4th order algorithm of
Yoshida~\cite{syoshida_4th_order}.  We employed a parallelogram,
non-primitive unit cell containing four Fermi potentials. The full
potential was represented as a sum over all Fermi potentials Eq.(1) in
the unit cell and in its next and next-nearest neighboring unit cells.
For the figures in the main part we performed high precision
computations with an ensemble size of $\ge 10^5$ trajectories that
guaranteed a vanishingly small standard error of the mean in Eq.(2).
The initial conditions were sampled uniformly in the coordinate space
of the unit cell, the initial speeds of the particles were fixed to
satisfy the total energy condition ($E=1/2$), and the initial launch
angles were randomized.  A numerical estimate for the diffusion
coefficient $D$ Eq.(2) was obtained with a linear fit to
$\langle(\mathbf{r}(t)-\mathbf{r}(0))^2\rangle$ as a function of time,
where we skipped the initial transient region and instead made a fit
at $t=1000\ldots 5000$. The two figures shown later in this Supplement
were generated with less precision than in the main part. In both
cases we iterated up to time $t=5000$ with a time step of $10^{-3}$.
For Fig.~\ref{fig:dsigma} we chose an ensemble size of $10000$
particles, for Fig.~\ref{fig:dws} we had $100000$ particles.

\section{2. Random walk approximations for diffusion}

In order to understand the coarse scale parameter dependence of the
diffusion coefficient $D(w)$ depicted in Fig.~2 we employ a
Boltzmann-type random walk approximation, which was put forward in
Ref.~\cite{sKlDe00}, see Sec.~4 therein. As briefly described in the
main text, this approximation is based on the assumption that
diffusion is governed by `flights' of length $\ell_c=\ell_c(w)$ over
corresponding flight time intervals $\tau_c=\tau_c(w)$ after which a
particle experiences a `collision'.  However, in contrast to the
standard Lorentz gas with hard walls studied in \cite{sKlDe00} our
potential is soft. Hence, here we define a collision as an event where
a particle hits the contour line of a scatterer at $E=1/2$ in the
triangular unit cell $A$ displayed in Fig.~1 of the main text. By
assuming that all collisions are uncorrelated the diffusion
coefficient can be approximated as 
\be 
D_B(w)=\frac{\ell_c^2}{4\tau_c}
\quad .
\label{eq:db}
\ee 
In the hard Lorentz gas $\tau_c$ can be calculated using a phase space
argument analogous to the one that was put forward in
Ref.~\cite{sMaZw83} to compute the mean escape time $\tau_e$ of a
particle out of a unit cell.  The latter can be expressed in terms of
the probability to leave a trap within the time $\tau_e$. This
quantity is in turn given by the portion of the phase space from which
a particle can escape from a trap during time $\tau_e$ divided by the
total phase space volume of a trap; for details we refer to
\cite{sMaZw83}. The only difference for computing $\tau_c$ is that here
one replaces the flux across the exits of a trap by the flux to
the walls bounding the trap. Working this out for our soft Lorentz gas 
we get
\begin{equation} 
  \tau^{-1}_c=\frac{v}{A} \quad .
\label{eq:tauc}
\end{equation}
Here $A=A(w)$ is the accessible area for the particle in position
space, and $v=|\mathbf{v}(w)|$ defines the average constant speed with
which a particle travels between collisions.  We now have everything
at hand to boil down Eq.(\ref{eq:db}) to something computable:
First, using $l_c=v\tau_c$ in Eq.(\ref{eq:db}) we trivially obtain \be
D_B(w)=\frac{v^2\tau_c}{4} \quad .  \ee Substituting $\tau_c$ by
Eq.(\ref{eq:tauc}) yields
\begin{equation}
D_B(w)=\frac{A}{4}v \quad .
\end{equation}
$A$ is easily computed by geometric arguments leading to
our central formula
\begin{equation}
D_B(w)=\frac{\sqrt3L^2/4-\pi r_0^2/2}{4}v \quad ,
\label{eq:dbf}
\end{equation}
where $L$ is the lattice spacing. It remains to calculate $v$. For
this recall that a particle moves in the plane under the influence of
overlapping Fermi potentials, see Eq.(1) in the main text. This means
the kinetic energy varies depending on the position of the particle,
consequently the speed fluctuates as well. However, as explained
above, for Eq.(\ref{eq:dbf}) we assume that a particle travels with an
on average constant speed $v$.  We define this speed in two ways by
using the following approximations:
\begin{enumerate}

\item We calculate {\em analytically} an {\em approximate average
  speed $v_\mathrm{ave}=v_\mathrm{ave}(w)$} at the moment when a
  particle leaves $A$. For this we consider the contributions of the
  potential from two adjacent lattice points in the plane only.
  Without loss of generality we may choose $(0,0)$ and $(L,0)$ located
  at the centres of two nearby potentials $V_1(x):=V_1(x,0)$ and
  $V_2(x):=V_2(L,0)$ with $L=2r_0+w$. Note that with the latter
  equation for the lattice spacing we approximate the true gap size in
  the case of overlapping Fermi potentials by a gap size $w$ derived
  from using a single non-overlapping Fermi potential; see
  Ref.~\cite{sGall18} for details.  By considering the contributions
  from these two potentials along the $x$-axis the joint potential
  $V_j(x)$ reads
\begin{equation}
  V_j(x)=V_1(x)+V_2(x)=\frac{1}{1+\exp((|x|-r_0)/\sigma)}+\frac{1}{1+\exp(|x-L|-r_0)/\sigma)} \quad .
\label{V1plusV21d}
\end{equation}
We now define an average potential $V_{\rm ave}$ over the approximate exit of
a trap according to
\begin{equation}
V_{\rm ave}(w)=\frac{1}{w}\int_{r_0}^{r_0+w}V_j(x)dx \quad .
\end{equation}
Exploiting symmetry this integral can be calculated to
\begin{equation}
V_{\rm ave}(w)=2+\frac{2\sigma}{w}\ln\bigg(\dfrac{2}{1+\exp(w/\sigma)}\bigg)\quad.
\label{v_ave}
\end{equation}
Conservation of energy yields $v=\sqrt{2(E-V_j(x))}$. Combining this with Eq.~(\ref{v_ave}), an average exit speed can be expressed as 
\begin{equation}
v_{\rm ave}(w)=\sqrt{2(E-V_{\rm ave}(w))}\quad .
\label{velave}
\end{equation}

\item A second definition is based on calculating {\em numerically} the {\em correct average
  speed} $v_\mathrm{num}=v_\mathrm{num}(w)$ while a particle is leaving
  a trap. Using symmetry we have to compute the integral
\be
v_\mathrm{num} = \frac{2}{w} \int\limits_{r_\epsilon}^{r_{\epsilon}+w/2} \sqrt{2\left(E-V_\mathrm{tot}(x)\right)}\,\mathrm{d}x,
\ee
where $V_\mathrm{tot}(x):=V_\mathrm{tot}(x,0)$ is the sum over the
range of potentials as defined in Sec.~I above. Note that this
requires us to compute $r_\epsilon$ defined by
$V_\mathrm{tot}(r_\epsilon) = 1/2$ due to the overlapping potentials.
This integral is not solvable analytically, hence we compute it
numerically.
\end{enumerate}

Using these two approximations for the speed $v$ in Eq.(\ref{eq:dbf})
yields our two approximations $D_B$ and $D_{B,num}$ plotted in Fig.~2
of the main text.

\newpage

\section{3. Transition between the soft and the hard Lorentz gas}

In this section we discuss first qualitatively and then quantitatively
the changes of the diffusive dynamics in the hard disk Lorentz gas
when softening the scatterers. In Subsec.~A we outline generic
dynamical systems properties of both models by discussing their
similarities and differences. Subsection~B presents computer
simulation results for the diffusion coefficient as a function of the
smoothness parameter $\sigma$ exploring this transition in more depth
when $\sigma$ is getting smaller.

\subsection{3.1. The hard and the soft Lorentz gas: Similarities and
  differences in their diffusive dynamics}

Diffusion in the conventional Lorentz gas consisting of a point
particle scattering with hard disks has been studied in a large number
of works; see \cite{sGasp,sDo99,sKla06,sSza00,sDett14} for reviews.
Pioneering mathematical research by Bunimovich and Sinai showed
rigorously that the Lorentz gas is a K-system, which implies that it
is mixing and ergodic \cite{sBuSi80a,sBuSi80b}. It is furthermore a
hyperbolic chaotic dynamical system \cite{sSin70}. For the
two-dimensional periodic Lorentz gas with scatterers situated on a
triangular lattice these strong chaos properties imply that diffusion
is normal in the parameter regime of the gap size $w$ of
$0<w<w_{\infty}$, in the sense that the mean square displacement (MSD)
grows linearly with time in the long-time limit, as was proven in
Ref.~\cite{sBuSi80a}.  In this regime the diffusion coefficient as a
function of $w$ was explored in
Refs.~\cite{sMaZw83,sKlDe00,sKlKo02,sGilSan09,sAnMo12}. The result from
simulations of $D(w)$ is shown in the inset of Fig.~2 in the main
text. While in this plot $D$ looks like a rather smooth function of
$w$ it was shown in Ref.~\cite{sKlDe00} that there are irregularities
in the form of slight wiggles on very fine scales.  The order of
magnitude of these irregularities is by far smaller than in the soft
model, cp.\ Fig.~2 in the main text for the soft Lorentz gas with
Fig.~1(b) in \cite{sKlDe00}. So far it is only known that these
irregularities exist in the conventional Lorentz gas, but they could
not be explained, e.g., in terms of periodic orbits.  There are
attempts in the literature, however, to compute $D(w)$ for the hard
periodic Lorentz gas in terms of {\em unstable} periodic orbits
\cite{sCvGS92,sCEG91,sCAMTV01}.

Note that $w_{\infty}=4\sqrt{3}-3\simeq0.3094$ defines the onset of an
{\em infinite horizon} in the periodic Lorentz gas, where a particle
can for the first time move ballistically along infinite channels
across the entire lattice without colliding with any scatterer.
Accordingly, for $w_{\infty}> w$ diffusion becomes anomalous, and $D$
as defined by Eq.~(2) in the main text does not exist anymore.  This
is reflected in our chart of periodic orbits in Fig.~4 (main part) as
the big red tongue of quasi-ballistic periodic orbits emerging from
$w\simeq 0.31$ for small $\sigma$. More precisely, infinite horizon
Lorentz gases exhibit a special type of superdiffusion, where the MSD
grows like $t\ln t$. This is due to a family of strictly ballistic
periodic orbits which, however, occupy only zero volume in the whole
phase space of the system
\cite{sZGNR86,sFVG17,sDett12,sCGLS14,sCGLS14c,sZPFDB18}.

In marked contrast to this hard Lorentz gas scenario of purely chaotic
deterministic diffusion with a well-defined diffusion coefficient for
$w<w_{\infty}$, and superdiffusion for $w>w_{\infty}$ due to an
infinite horizon, in our main text we have shown that softening the
hard disks by using overlapping Fermi potentials changes the diffusive
dynamics profoundly: Even a minimal softening of the potential yields
a mixed phase space \cite{sLiLi92} composed of chaotic regions
interrupted by small `islands' of periodic orbits, cf.\ Fig.~3 (main
part). Note that in contrast to the hard Lorentz gas these islands
occupy non-zero volume and are {\em stable} in phase space. This
implies a completely different type of non-hyperbolic dynamics
compared to the well-behaved hard Lorentz gas which, in turn, is
reflected in profound changes of the corresponding diffusive
properties. In turn, it is well-known that islands of periodicity in
phase space corresponding to `propagating' periodic orbits, also
called accelerator modes, lead to superdiffusion, even if they are
very tiny, while parameter regions without any islands correspond to
normal diffusion \cite{sGZR87,sGZR88,sZas02,sKRS08,sMaRo14,sHaCo18}.

At first view this phenomenon may look similar to the infinite horizon
case in the hard periodic Lorentz gas. But in the soft model the
orbits generating superdiffusion are not strictly `ballistic' in the
sense of being collision-free, as is demonstrated by Fig.~3 (main
text). Rather particles collide with the scatterers in intricate ways
while they move on average in one direction. Hence there are no
infinite horizon channels in our soft system, as there is always a
force acting on a particle, therefore we call these trajectories {\em
  quasi-ballistic}. Secondly, in contrast to the hard Lorentz gas the
phase space of a quasi-ballistic island of periodicity is non-zero,
due to the velocity now representing an additional degree of freedom.
This implies a generically different type of superdiffusion compared
to the hard Lorentz gas infinite horizon case with a MSD that grows
like $t^{\alpha}\:,\:\alpha>1$
\cite{sZGNR86,sFVG17,sDett12,sCGLS14,sCGLS14c,sZPFDB18}.

To our knowledge so far there are no works that study the transition
from Hamiltonian particle billiards consisting of hard walls, like
Lorentz gases, to systems consisting of soft periodic potentials, like
our model, as far as diffusion is concerned. The only research we are
aware of is a series of articles by Turaev and Rom-Kedar, who
investigate in mathematical depth the changes of the dynamics by
softening the hard walls of billiards, however, without exploring the
impact on diffusive properties \cite{sTRK98,sRKT99}.  There is a claim
in these references that under softening hard walls islands of
periodicity become typical in parameter space, but this has not been
observed in parameter space as shown in our Fig.~4. We also remark
that the theory by Turaev and Rom-Kedar predicts that for small
$\sigma$ the tongues of periodic orbits depicted in Fig.~4 should grow
linearly in parameter space, see Theorem~1 in Ref.~\cite{sRKT99}.  But
to verify this numerically is at present out of reach, as it is
extremely difficult to find small islands of periodicity especially
for $\sigma$ close to zero.

\subsection{3.2. The diffusion coefficient as a function of the smoothness
  parameter}

We now explore the transition between the hard and the soft Lorentz
gas regarding their diffusive properties in more depth. In the main
text we have focused on the diffusion coefficient as a function of the
gap size $w$ between two adjacent scatterers for fixed smoothness
parameter $\sigma$, cf.\ Fig.~2. Here we first present results for the
diffusion coefficient $D$ at fixed $w$ under variation of $\sigma$.
This supplements our previous analysis related to the chart of
periodic orbits shown in Fig.~4, main part: While so far we have
explored the parameter space displayed therein along horizontal cuts
through this plane, in the following we elucidate what happens along
vertical cuts. This sheds light on the transition of diffusive
properties between the soft and the hard scatterer case when
$\sigma$ is close to zero.

\begin{figure}[bh!]
\centering
\begin{minipage}[c]{0.49\textwidth}
\centering \includegraphics[width=0.99\textwidth]{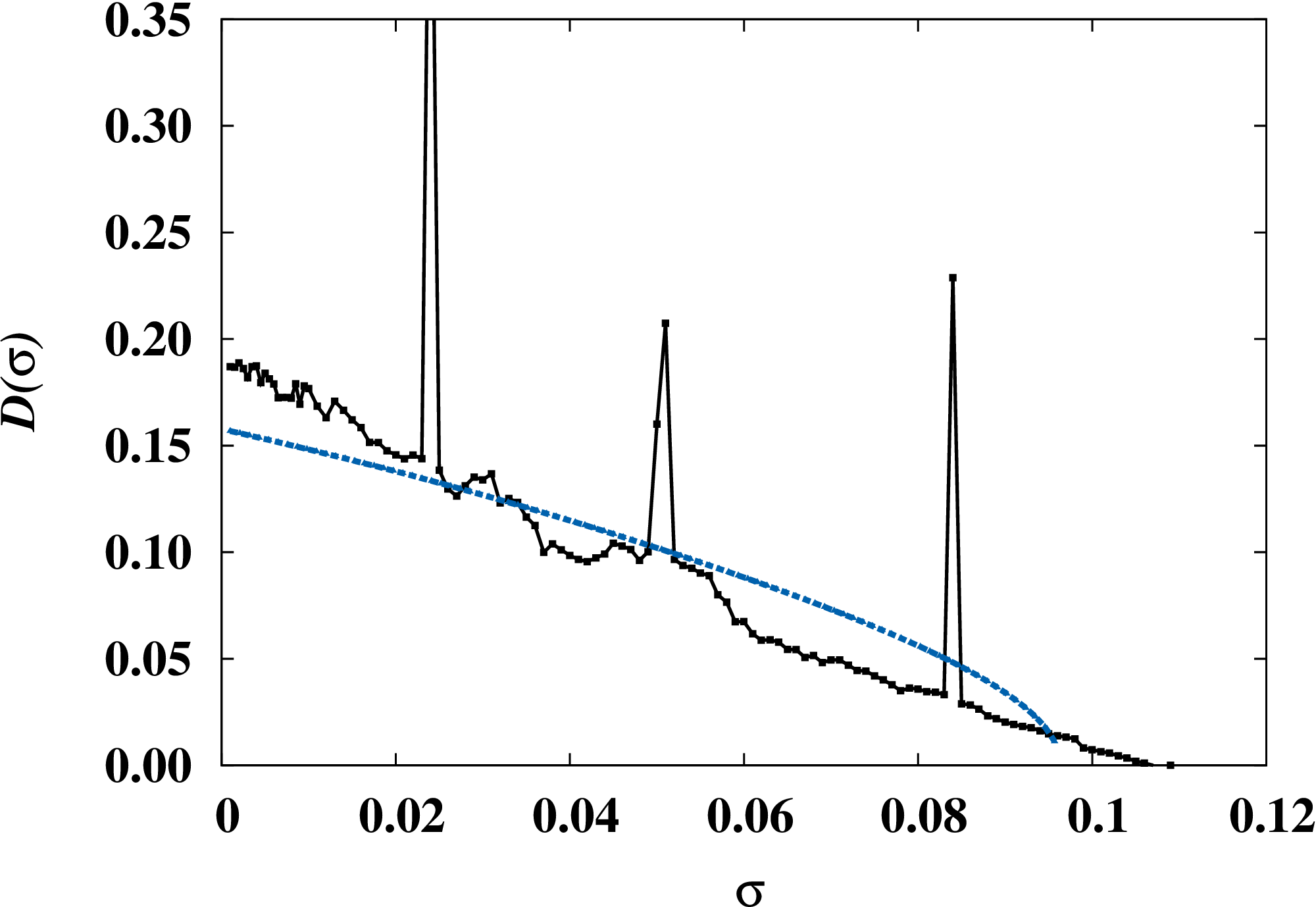}
\end{minipage}
\begin{minipage}[c]{0.49\textwidth}
\centering \includegraphics[width=0.99\textwidth]{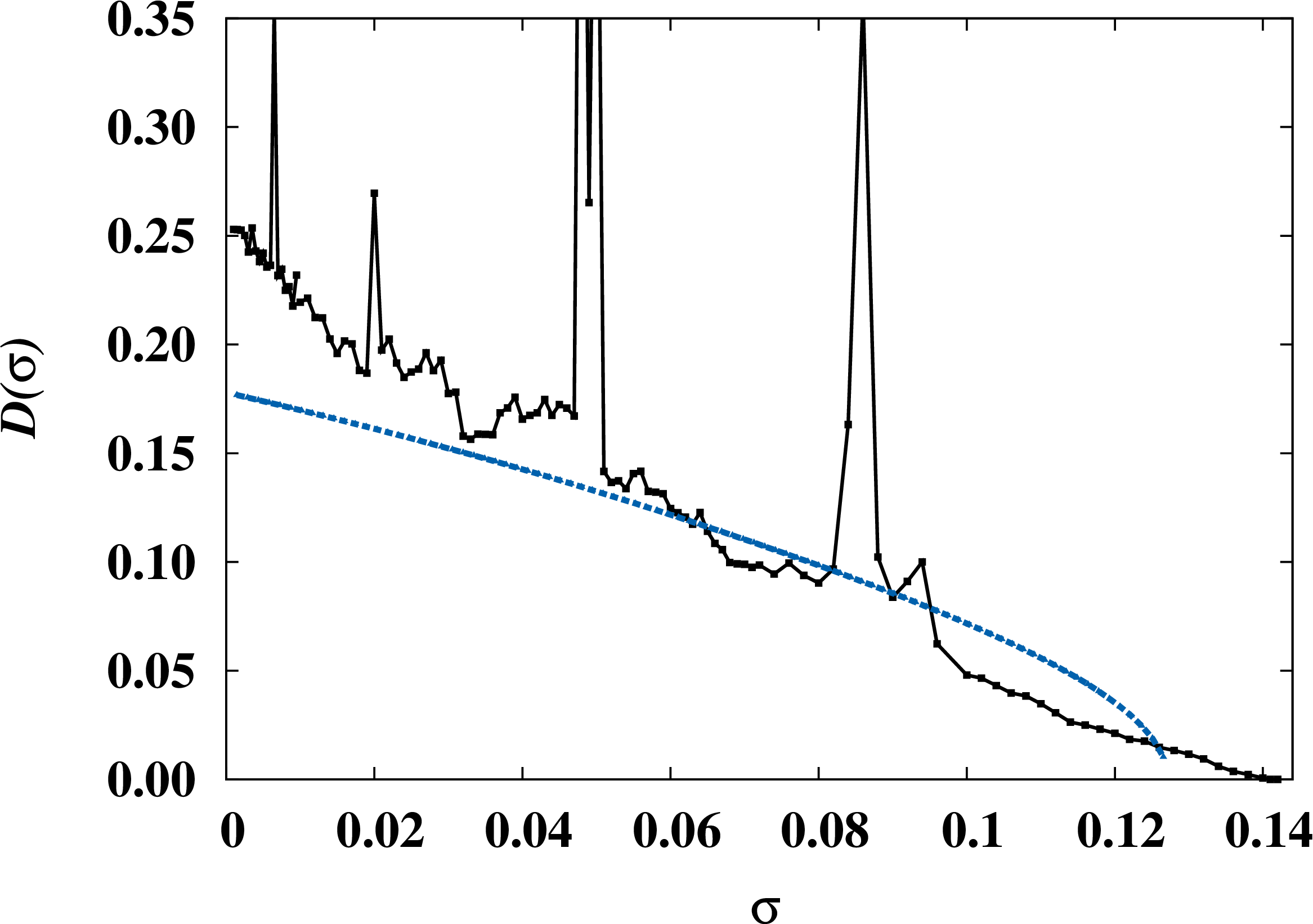}
\end{minipage}
\caption{Diffusion coefficient $D$ as a function of the smoothness
  parameter $\sigma$ when the gap size is fixed to $w=0.235$ (left)
  and $w=0.31$ (right). The black lines with symbols represent results
  from computer simulations. The peaks therein correspond to parameter
  regions where the diffusion coefficient has not converged to a final
  value due to the existence of quasi-ballistic periodic orbits. The
  (blue) dashed lines give the simple random walk approximation
  $D_{\rm MZ}(\sigma)$ Eq.~(\ref{eq:dsmz}).}
\label{fig:dsigma}
\end{figure}

Figure~\ref{fig:dsigma} (a) and (b) depict results for $D(\sigma)$ at
two fixed values of $w$. These two figures correspond to vertical cuts
in Fig.~4 (main text) from top to bottom by showing what happens when
$\sigma$ approaches zero. We see that in both cases $D(\sigma)$ is an
increasing function when decreasing $\sigma$. However, as in case of
$D(w)$ for fixed $\sigma$ in Fig.~2 (main text), we observe again
irregularities on finer scales. They are supplemented by peaks
representing parameter regions where $D$ has not converged to a final
value, as it does not exist due to the existence of quasi-ballistic
periodic orbits. These peaks match to respective quasi-ballistic
tongues in Fig.~4 (main text).  Note, however, that in the left figure
the lowest tongue around $\sigma\simeq0.01$ has been missed. This may
be due to the chosen spacing between two adjacent data points
$D(\sigma)$, or that our initial ensemble did not catch a respective
tiny island of stability. The values of $D(0)$ correspond to the
respective results for the hard Lorentz gas shown in the inset of
Fig.~2 (main text).

The blue dashed lines are analytical results representing a simple
random walk approximation put forward by Machta and Zwanzig
\cite{sMaZw83}, suitably adapted to the soft Lorentz gas: The
assumption is that particles hop from trap to trap on a triangular
lattice, cf.\ Fig.~1 (main text) for the trapping region $A$ with an
escape time $\tau_e$ from each trap, where the centres of the traps
are a distance $\ell=L/\sqrt{3}$ apart. By assuming no memory between
two jumps, similarly to Eq.~(\ref{eq:db}) above yielding our Boltzmann
approximation the diffusion coefficient can be approximated to
\begin{equation}
D_{\rm MZ}(\sigma)=\frac{\ell^2}{4\tau_e} \quad .
\label{eq:dsmz}
\end{equation}
Again in analogy to the Boltzmann approximation $\tau_e$ is now given
again by a phase space argument,
\begin{equation}
\tau_e^{-1} =\frac{3wv}{\pi A}\quad ,
\label{eq:taue}
\end{equation}
where, as explained in Sec.~\ref{sec:numsim} in the Supplement, the
only difference to the collision time $\tau_c$ in Eq.~(\ref{eq:tauc})
is that for calculating the escape time $\tau_e$ we consider the
portion of phase space where a particle leaves a trap. Combining
Eq.~(\ref{eq:taue}) with Eq.~(\ref{eq:dsmz}) by plugging in the value
for $A$ as before yields
\begin{equation}
D_{\rm MZ}(\sigma)=\frac{L^2w}{\pi(\sqrt{3}L^2-2\pi)}v \quad .
\label{Dif_w_v}
\end{equation}
In Fig.~\ref{fig:dsigma} above we have used this formula by replacing
$v$ with the average velocity calculated in
Eqs.~(\ref{v_ave}),(\ref{velave}) above, $v=v_{ave}$, which yields the
dashed blue lines. We see that this analytical random walk
approximation matches qualitatively to the numerical results by
particularly explaining the increase of the (normal) diffusion
coefficient when the system approaches the hard Lorentz gas limit for
small $\sigma$. Note that the quantitative mismatch between the data
and the approximation is increasing from $\sigma\to0$, as is analysed
in detail in \cite{sKlDe00,sKlKo02}. What we can learn from
Fig.~\ref{fig:dsigma} and its analysis is that the transition between
diffusion in the soft and the hard Lorentz gas for $\sigma$ close to
zero looks smooth on a coarse grained level as far as the diffusion
coefficient is concerned when it exists, as is confirmed by our random
walk approximation. However, whenever quasi-ballistic islands of
periodicity occur, they interrupt this scenario.  Increasing the
numerical precision will reveal more and more superdiffusive parameter
regions, probably even an infinite set of them, thus severely
disrupting any smooth transition scenario.  This result is fully in
line with our chart of periodic orbits Fig.~4 (main text) by
illustrating it in detail for $D(\sigma)$.

\begin{figure}[t]
	\begin{center}		
	\includegraphics[angle=-90,width=0.75\textwidth]{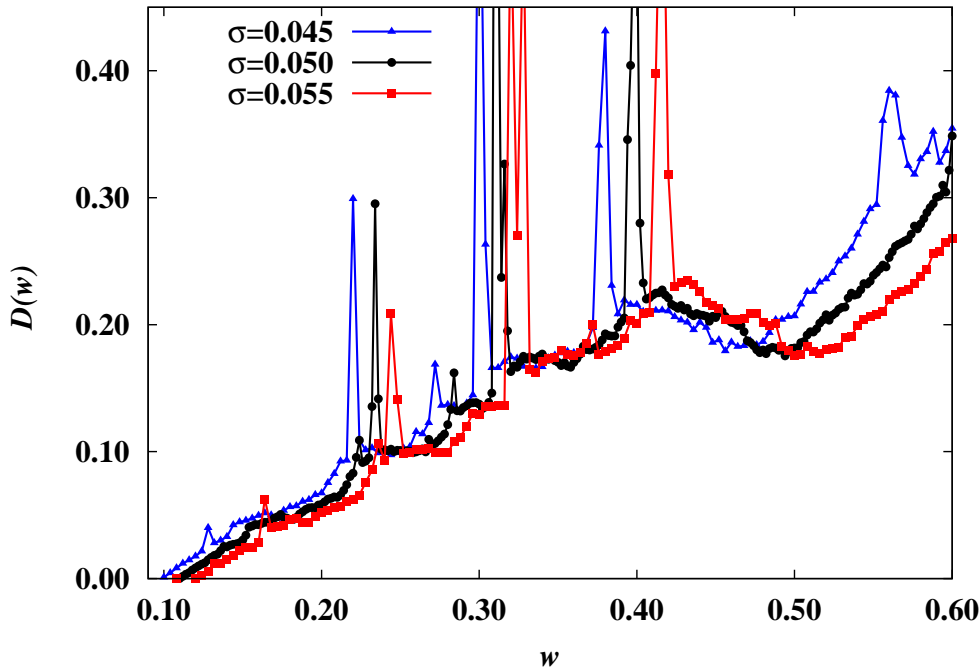}
	\end{center}
	\caption{Diffusion coefficient $D(w)$ obtained from
          simulations for three different values of the smoothness
          parameter $\sigma$ as given in the figure.}
	\label{fig:dws}
\end{figure}

We finish our discussion of the impact of the smoothness parameter on
diffusion in the soft Lorentz gas by presenting numerical results for
$D(w)$ at three different values of $\sigma$, see Fig.~\ref{fig:dws}.
The shift of the different peaks, where diffusion is anomalous, to the
left when $\sigma$ is getting smaller is again fully in line with our
chart of periodic orbits Fig.~4 (main text). We see that the whole
curve where the normal diffusion coefficient exists is slightly
deforming under variation of $\sigma$: Except in a tiny parameter
region of $0.4<w<0.5$ overall it is increasing when $\sigma$ is
getting smaller, i.e., when the soft system is approaching the hard
Lorentz gas limit. Within the parameter region of $0<w<0.31$
eventually it will converge to the known diffusion coefficient of the
hard Lorentz shown in the inset of Fig.~2 (main text) while the whole
parameter region for $w>0.31$ will gradually become superdiffusive.

\newpage



\end{appendix}

\end{document}